\documentclass[aps,prb,twocolumn,superscriptaddress,showpacs]{revtex4}
\usepackage{graphics}
\usepackage{graphicx}
\usepackage[english]{babel}
\usepackage{color}

\begin{document}

\title{Excess of topological defects induced by confinement in vortex nanocrystals}

%------------------------------------------------------------------------------------------------------------------------------------------------

\author{N. R. Cejas Bolecek}
%\email[]{ncejas@cab.cnea.gov.ar}
%%\homepage[]{Your web page}
%%\thanks{}
%%\altaffiliation{}
\affiliation{Laboratorio de Bajas Temperaturas, Centro At\'{o}mico
Bariloche \& Instituto Balseiro, Bariloche, Argentina}
\author{ M. I. Dolz}
\affiliation{Departamento de F\'isica, Universidad Nacional de San Luis, CONICET, Argentina}
\author{H. Pastoriza}
\affiliation{Laboratorio de Bajas Temperaturas, Centro At\'{o}mico
Bariloche \& Instituto Balseiro, Bariloche, Argentina}
\author{M. Konczykowski}
\affiliation{Laboratoire des Solides Irradi\'{e}s, \'{E}cole
Polytechnique, CNRS, CEA, Universit\'{e} Paris-Saclay, Palaiseau,
France}
\author{C. J. van der Beek}
\affiliation{Laboratoire des Solides Irradi\'{e}s, \'{E}cole
Polytechnique, CNRS, CEA, Universit\'{e} Paris-Saclay, Palaiseau,
France}
\author{ A. B. Kolton}
\affiliation{Teor\'{\i}a de la Materia Condensada, Centro
At\'{o}mico Bariloche \& Instituto Balseiro, Bariloche, Argentina}
\author{Y. Fasano\footnote{corresponding author:yanina.fasano@cab.cnea.gov.ar} }
\affiliation{Laboratorio de Bajas Temperaturas, Centro At\'{o}mico
Bariloche \& Instituto Balseiro, Bariloche, Argentina}
%------------------------------------------------------------------------------------------------------------------------------------------------

%------------------------------------------------------------------------------------------------------------------------------------------------

\date{\today}

\begin{abstract}

We directly image individual vortex positions in nanocrystals in
order to unveil the structural property that contributes to the
depletion of the entropy-jump entailed at the first-order
transition. On reducing the nanocrystal size the density of
topological defects increases near the edges over a characteristic
length. Within this ``healing-length'' distance from the sample edge
vortex rows tend to bend  while towards the center of the sample the
positional order of the  vortex structure is what is expected for
the Bragg-glass phase. This suggests that the healing-length may be
a key quantity to model the entropy-jump depletion in the
first-order transition of extremely-layered vortex nanocrystals.
\end{abstract}


\maketitle

\section{Introduction}

The growing demand of  miniaturization in superconducting devices
applied to the detection of different types of radiation
\cite{welp13,sekimoto13,kashiwagi14,kashiwagi14b, watanabe14} and
magnetic signals \cite{takeda08,yamaguchi10, schwarz12} has
triggered the study of thermodynamic and transport properties of
high-$T_{\rm c}$ superconducting materials at the micro and
nanoscale. Since many of these devices operate in the mixed state,
understanding the change in thermodynamic and structural properties
of vortex matter when reducing the number of vortices down to the
nanoscale is crucial for predicting their working-range. For
instance, transition-edge superconducting devices are based on the
detection of a sudden increase of dissipation due to particular
events when continuously measuring voltage with a low applied
current. If confinement changes the critical current up to which
vortices do not dissipate, then the detection edge needs to be
readjusted.

From a fundamental point of view, vortex matter in type-II
superconductivity is a case-study for understanding how the physical
properties and phase diagrams change when going from macroscopic to
nanocrystalline condensed matter. In the case of hard
condensed-matter, nanocrystals are made up of at most a few thousand
of particles (atoms);~\cite{Coombes1972,Goldstein1992} similarly
soft-condensed matter ``vortex nanocrystals'' can be nucleated in
micron-sized samples with the same amount of
vortices.~\cite{Geim1997,Geurts2007,Connoly2008,Chao2009,Cren2011,Pereira2011,Tominaga2012,Mills2015}
Typically, hard condensed matter nanocrystals present a decrease of
transition temperatures, entropy and enthalpy jumps in melting and
solid-solid first-order phase
transitions.~\cite{Coombes1972,Goldstein1992,Tolbert1994,Guisbiers2009}
This is the consequence of a depletion of the total binding energy
since the particle's surface-to-volume ratio  increases on
decreasing the system size. In the case of nanocrystalline vortex
matter confinement effects also affect the transition lines and
structural properties,~\cite{Schweigert1998,Bolecek2015} although
different degrees of freedom and interactions are at play.
Particularly, in the case of vortex nanocrystals nucleated in
extremely-layered high-temperature superconductors the phase diagram
is finely tuned by many energy scales.~\cite{HabKees} The interplay
among inter-vortex interactions, thermal fluctuations, pinning, and
extremely anisotropic magnetic properties can be controlled by
applied field, temperature, crystalline disorder, and oxygen-doping,
respectively.

Recently, some of us reported~\cite{Dolz2015} on the peculiarities
of decreasing the system size in extremely-layered vortex matter:
for roughly hundred particles (vortices) no melting-point depression
is observed in contrast to results in hard condensed
matter.~\cite{Coombes1972} The entropy-jump entailed at the
first-order transition decreases on reducing the system size and we
suggested that this might have origin in two effects that can
eventually occur simultaneously at the transition. First, since
there is evidence that the first-order transition might be
concomitant with a c-axis decoupling of pancake
vortices,~\cite{Pastoriza1994} confinement can induce an extra
contribution to decoupling and thus reduce the entropy-jump. Second,
confinement can also entail a deterioration  of the in-plane
structural order when nucleating vortex
nanocrystals.~\cite{Dolz2015} Unveiling the structural properties of
vortex nanocrystals is therefore mandatory in order to gain insight
and model the origin of the entropy-jump depletion.

In this work we study the evolution of the structural properties on
reducing the system size of nanocrystalline vortex matter. We
characterize the variation and spatial distribution of elastic and
plastic deformations in the quenched nanocrystalline vortex solid.
Our vortex nanocrystals with less than 4000 vortices are nucleated
in the whole area of micron-sized engineered samples at low applied
fields. We have direct access to the static structural properties
with single-vortex resolution.~\cite{Fasano1999} We present a
systematic study as a function of the number of vortices in the
nanocrystal, tuned either by sample physical size or vortex density.

\section{Experimental}

Nanocrystalline vortex matter is nucleated in micron-sized
Bi$_{2}$Sr$_{2}$CaCu$_{2}$O$_{8+y}$ disks with diameters $d$ ranging
$30\,$-$\,50\mu$m and thicknesses between 1-2\,$\mu$m. Disks are
engineered from optimally-doped  crystals ($T_{\rm c}=90$\,K) by
combining optical-lithography and physical ion-milling
techniques.~\cite{Dolz2010} During the last step of the sample
fabrication process, thin free-standing and freshly-cleaved disks
are obtained. The disks are placed on the sample-holder with
micro-manipulators and carefully glued with conducting epoxy such
that the cleaved surfaces remain clean.

 As a
result of the final cleaving process, more than 90\,\% of the disks
present sub-micron steps at the surface. We avoid using these disks
for our study when possible. In a very few cases, see for instance
Fig.\,\ref{Figure2}, we considered disks with sub-micron steps that
divide the sample in terraces with one of them containing less than
$\sim 10$\,\% of vortices. In these cases we only consider for our
analysis the roughly $90$\,\% of vortices that are at the same
terrace.

The nanocrystalline vortex solid is directly imaged with
single-vortex resolution by magnetically-decorating vortex positions
at 4.2\,K after field-cooling the sample  at low fields from $T
> T_{\rm c}$, as described in Ref.\,\onlinecite{Fasano2003}.
The micron-sized samples were obtained from the macroscopic sample
studied in Ref.\,\onlinecite{Dolz2014}, and part of the fabricated
disks were used to investigate the phase diagram of mesoscopic
vortex matter reported in Ref.\,\onlinecite{Dolz2015}.

\section{Results}

\begin{figure*}
\includegraphics[width=\textwidth]{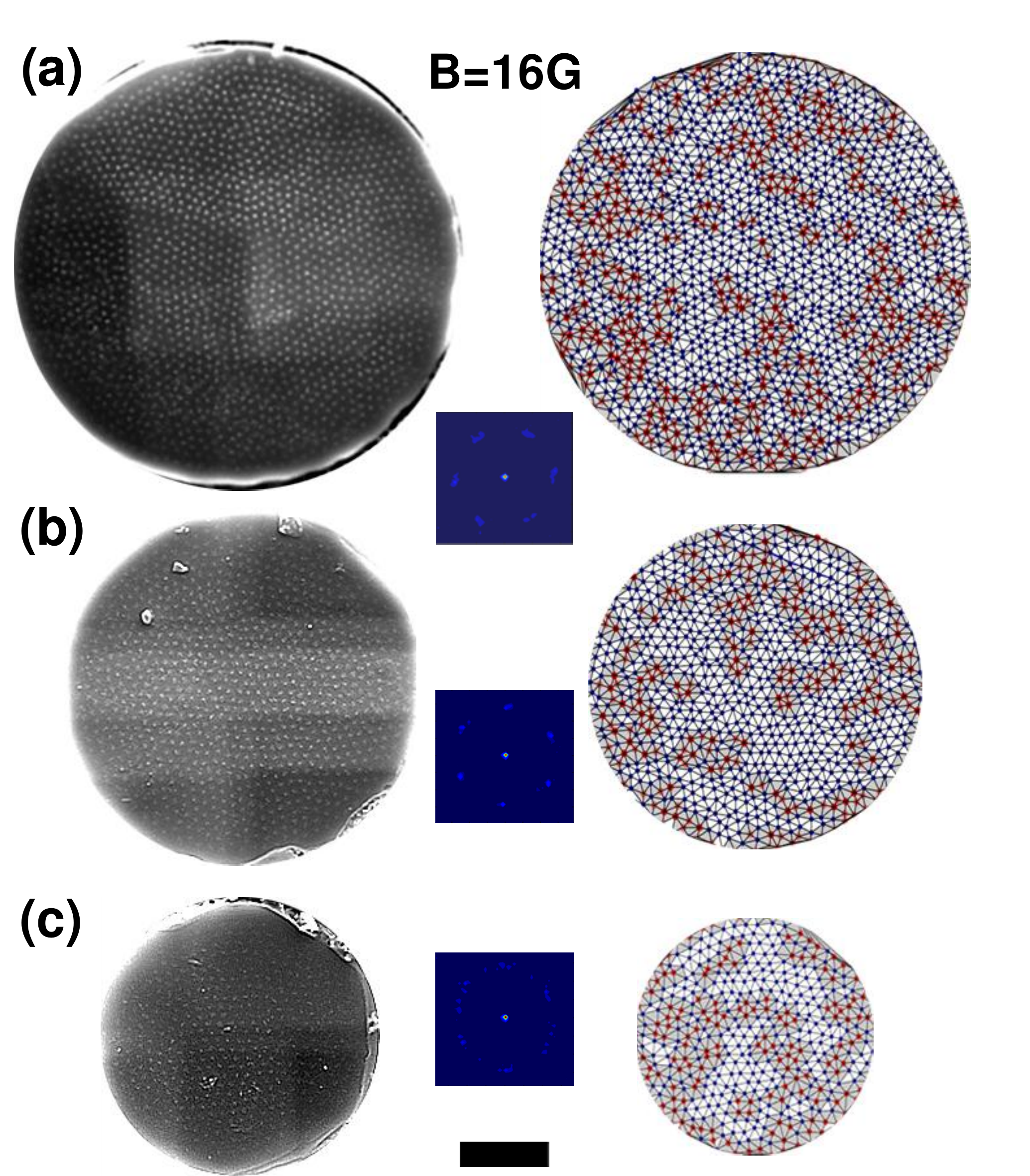}
\caption{\label{Figure1} Vortex nanocrystals with a vortex density
of 16\,G nucleated in micron-sized
Bi$_2$Sr$_2$CaCu$_2$O$_{8+\delta}$ disks with diameters of (a) 50,
(b) 40 and (c) 30\,$\mu$m. Left panels: vortices imaged in white by
means of field-cooling magnetic decorations performed at 4.2\,K.
Central panels: Fourier transforms of the vortex positions. Right
panels: Delaunay triangulations of the vortex structure depicting
non-sixfold (sixfold) coordinated vortices in red (blue) with
plastic deformations highlighted in gray. The scale-bar indicates
$10\,\mu$m.}
\end{figure*}

\begin{figure*}[ttt]
\includegraphics[width=\textwidth]{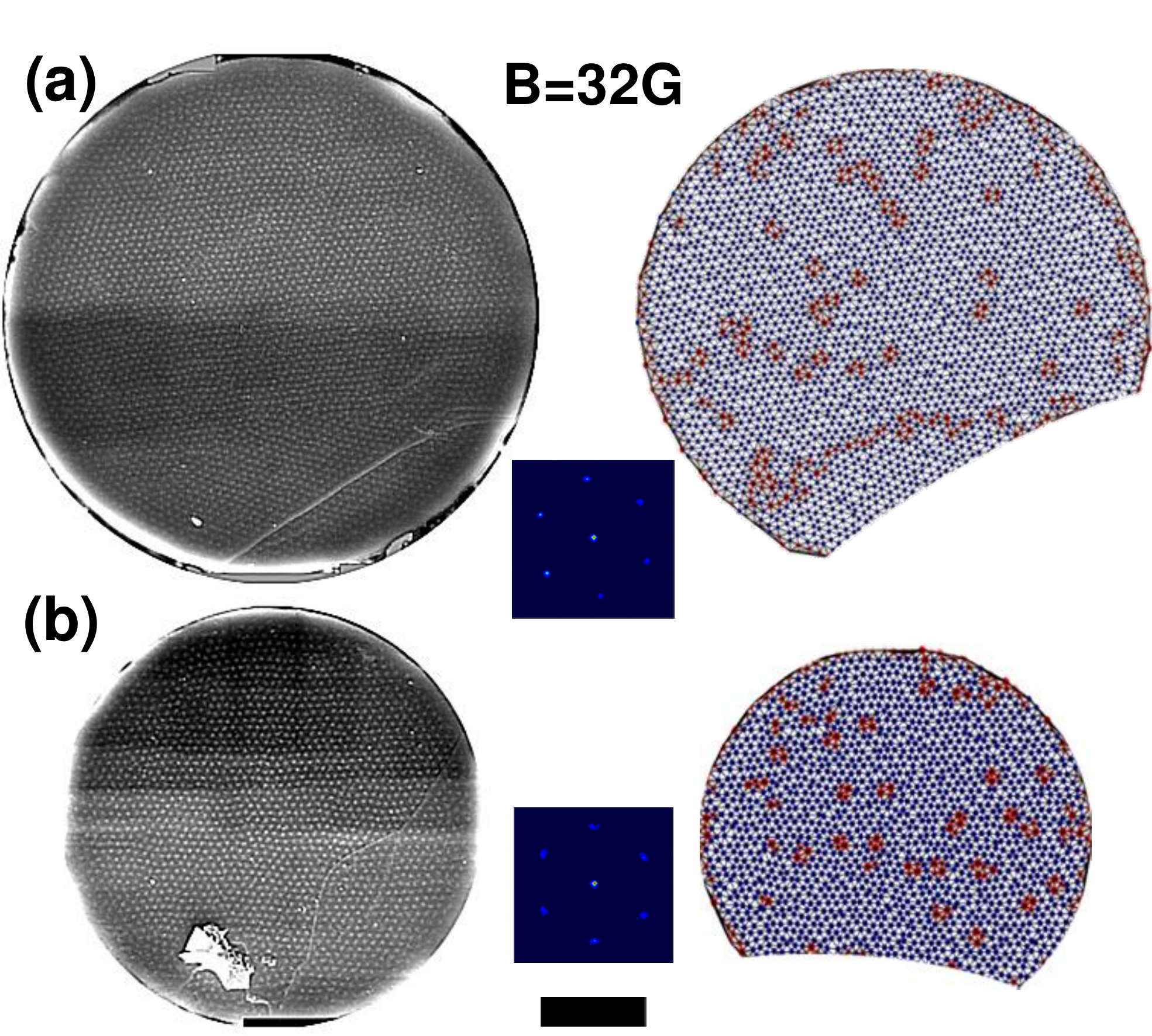}
\caption{\label{Figure2} Vortex nanocrystals with a vortex density
of 32\,G nucleated in micron-sized
Bi$_2$Sr$_2$CaCu$_2$O$_{8+\delta}$ disks with  diameters of  (a) 50
and  (b) 40\,$\mu$m. Left panels: vortices imaged  by means of
field-cooling magnetic decorations performed at 4.2\,K. Central
panels: Fourier transforms of vortex positions. Right panels:
Delaunay triangulations showing non-sixfold (sixfold) coordinated
vortices in red (blue) and  topological defects in gray. The
scale-bar indicates $10\,\mu$m.}
\end{figure*}

Figure \ref{Figure1} shows real-space images of vortex nanocrystals
with a vortex density of $16\,$G nucleated in micron-sized
Bi$_2$Sr$_2$CaCu$_2$O$_{8+\delta}$ disks with $d$ ranging 50 to
30\,$\mu$m (with approximately 1500 to 400 vortices). The
nanocrystals have the outer vortex compact lines slightly bent,
following the edges of the samples. In all cases, this effect is
produced without a detectable change in vortex density (within
$2\,\%$) in the whole vortex nanocrystal. Vortices at the center of
the sample form a crystallite with decreasing size on reducing
field. As observed in Fig. \ref{Figure1}, these two structural
properties are achieved by accumulating plastic deformations towards
the edges of the nanocrystals. This is evident in the Delaunay
triangulations \cite{Fasano2005} of the right panels that indicate
the non-sixfold coordinated vortices in red and the neighborhood of
plastic deformations highlighted in gray. These topological defects
are mainly unpaired screw dislocations, each formed by a five- and a
seven-fold coordinated vortex (disclinations). The Fourier
transforms of
 the central panels of Fig.\,\ref{Figure1}
show six diffraction peaks that broaden  on decreasing sample size.
For the smallest studied sample, these peaks split up due to the
nucleation of two crystallites of similar size with compact planes
with a very small misalignment (smaller than 5\,$^{\circ}$), see
Fig.\,\ref{Figure1}(c). Therefore plastic deformations producing
topological defects in the vortex nanocrystal proliferate on
increasing confinement effects for a fixed vortex density.

\begin{figure*}[ttt]
\includegraphics[width=\textwidth]{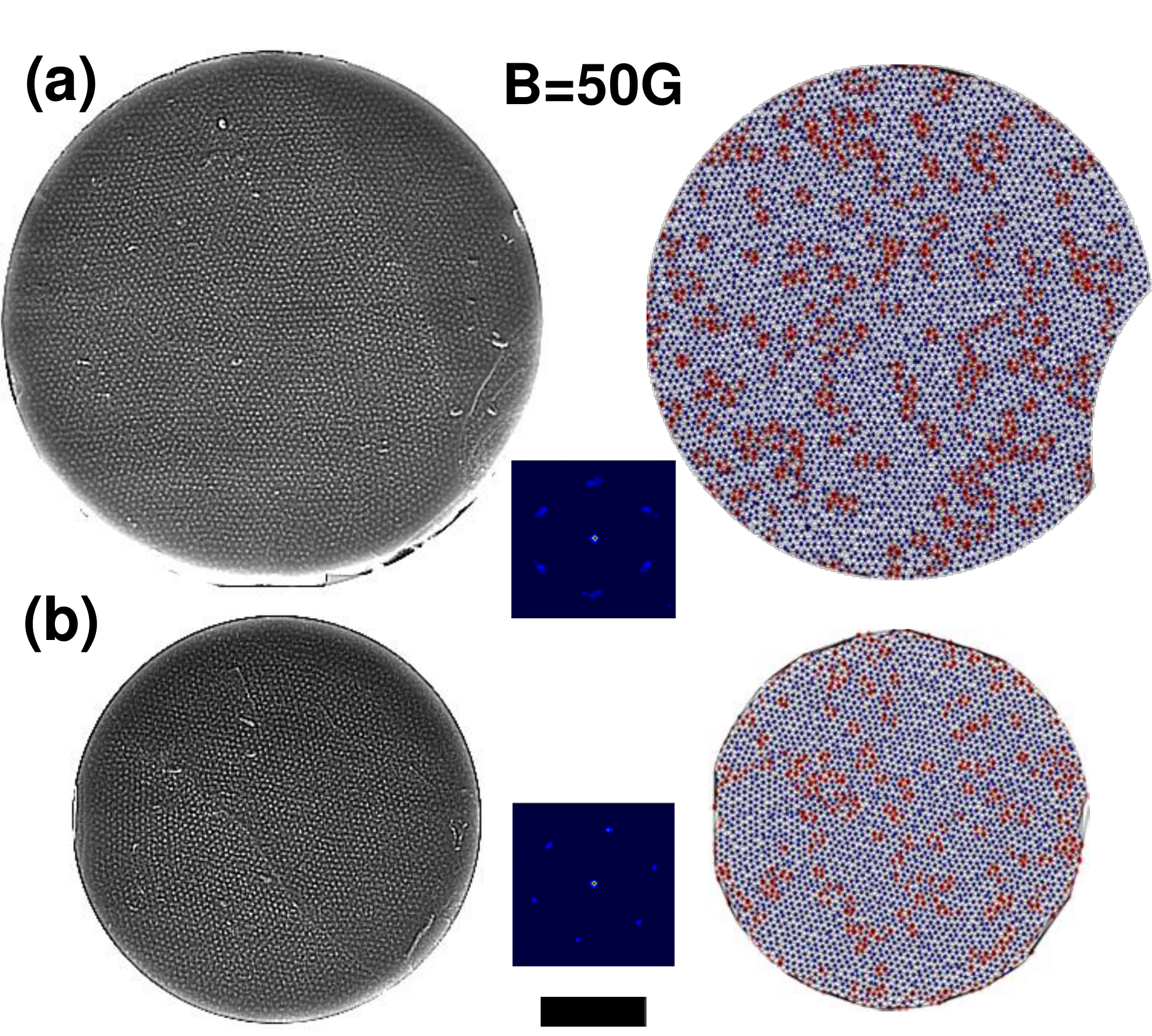}
\caption{\label{Figure3} Vortex nanocrystals with a vortex density
of 50\,G nucleated in micron-sized
Bi$_2$Sr$_2$CaCu$_2$O$_{8+\delta}$ disks with  diameters of  (a) 50
and  (b) 40\,$\mu$m. Left panels: vortices imaged  by field-cooling
magnetic decorations experiments performed at 4.2\,K. Central
panels: Fourier transforms of the vortex positions. Right panels:
Delaunay triangulations of the vortex structure depicting non
sixfold (sixfold) coordinated vortices in red (blue) with
topological defects in gray.  The scale-bar indicates $10\,\mu$m.}
\end{figure*}

Figures \ref{Figure2} and \ref{Figure3} show typical  nanocrystals
nucleated in 50 and 40\,$\mu$m diameter disks when the vortex
density  is increased roughly two and three times that of
Fig.\,\ref{Figure1}. Also in these cases the outer vortex compact
lines of the nanocrystals mimic the edges of the sample without any
noticeable local change in vortex density. For the higher vortex
densities of 32 and 50\,G the nanocrystal is formed by a single
crystal in the whole sample in contrast to the central crystallites
observed at 16\,G.  A particular case is that of the 32\,G vortex
structure nucleated in a 50\,$\mu$m diameter disk presenting a
sub-micron step at the sample surface. This feature induces a local
ordering of the vortex structure presenting one compact plane
parallel to the step in a region of less than 10\,$a$. As a result,
a planar grain boundary of paired screw dislocations is formed in
the nanocrystal. Besides this spurious effect,  on increasing the
nanocrystal stiffness ($B$) the  stress induced by the outer vortex
shells mimicking the sample edge produces a proliferation of
isolated clusters of topological defects. The size of the clusters
apparently increases with stiffness, namely $B$.

 We quantify the proliferation of topological defects  with the radial
density of topological defects, $\rho_{\rm def}(r)= N_{\rm def}(r) /
N_{\rm v} (r)$. We actually consider $N_{\rm def}$ y $N_{\rm v}$ at
a radius $r$ as the number of defects and of vortices included in a
circular shell with inner radius $r- \delta r/2,$ and outer radius
$r + \delta r/2$. The calculations are performed considering
concentric circular shells of width $\delta r = 2a$ and taking the
origin at the center of the vortex nanocrystal. The top panel of
Fig.\,4 shows the spatial distribution of defects in the considered
circular shells. Figure 4\,(a) shows $\rho_{\rm def}(r)$ at a fixed
vortex density of 16\,G for samples with three different radii,
whereas Fig.\,4\,(b) shows the results at densities of 16, 32 and
50\,G for the same sample size of $50$\,$\mu$m. The error bars in
the data correspond to the standard deviation of the values obtained
in different magnetic decorations for the same sample diameter and
$B$.

\begin{figure*}[ttt]
%\begin{center}
\includegraphics[width=\textwidth]{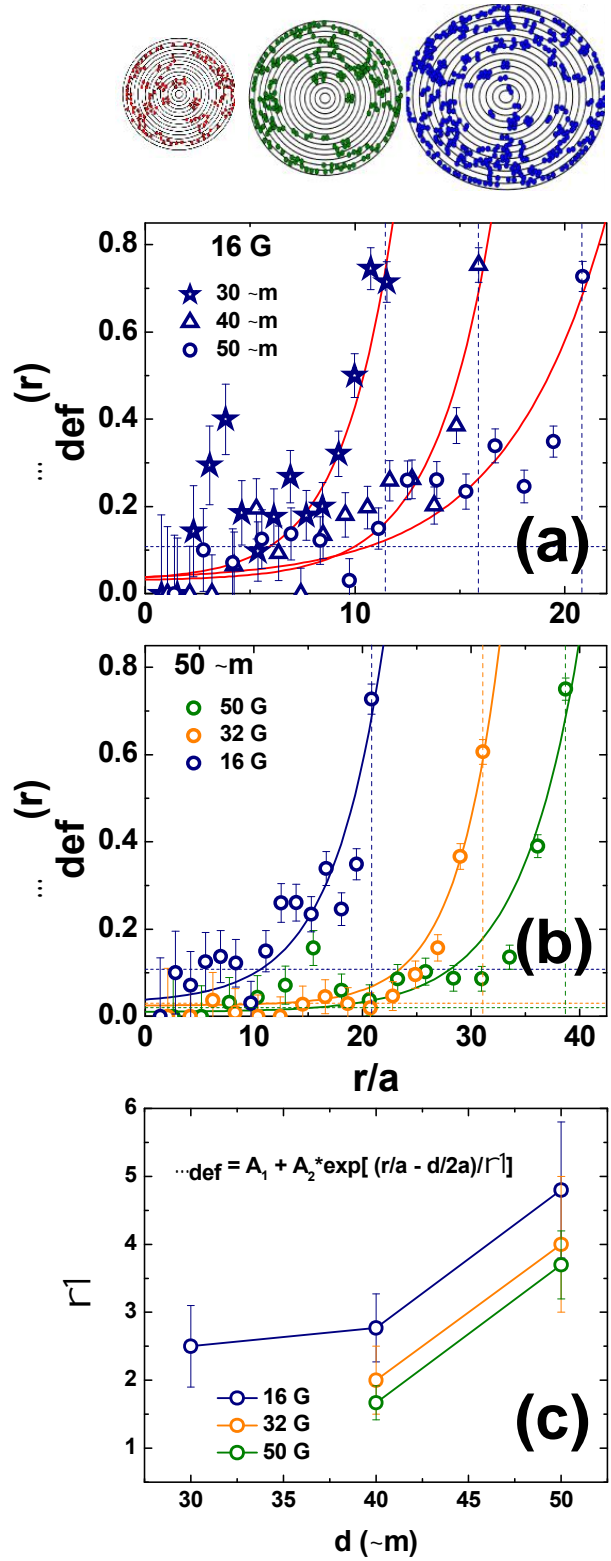}
%\end{center}
\caption{\label{Figure4} Radial density  of topological defects
$\rho_{\rm def}(r)$ as a function of  $r/a$ for nanocrystalline
Bi$_2$Sr$_2$CaCu$_2$O$_{8+\delta}$ vortex matter. (a) Vortex
nanocrystals with a density of 16\,G nucleated in 30, 40 and
50\,$\mu$m diameter disks. Top panel: circular shells considered to
calculate this magnitude and location of topological defects
(indicated with dots). (b) Vortex nanocrystals with densities of 16,
32 and 50\,G nucleated in 50\,$\mu$m diameter disks.  Vertical
dashed lines indicate the sample edge, $d/2$, and horizontal lines
the topological defects density for the parent macroscopic sample,
$\rho_{\rm def}^{\rm macro}$. The full lines are fits to the data
with a function $\rho_{\rm def}(r)= A_{1} + A_{2}\exp{[(r/a -
d/2a)/\alpha]}$. (c) Evolution of the healing length of vortex
nanocrystals, $\alpha$, with the sample diameter $d$ for the three
studied vortex densities.}
\end{figure*}

In all cases $\rho_{\rm def}(r)$ increases dramatically on
approaching the sample edge (indicated in the figures with dashed
lines) and stagnates on the center of the nanocrystals. These
saturation values roughly approach, within the dispersion of values
for different experimental realizations,  the density of defects
found in macroscopic samples for each $B$, $\rho_{\rm def}^{\rm
macro}=N_{\rm def}^{\rm macro}/N^{\rm macro}_{\rm v}$. For the
macroscopic parent sample from which the disks were engineered,
$\rho_{\rm def}^{\rm macro} \sim 11$, 3 and 2\,\% for 16, 32 and
50\,G structures, respectively \cite{Bolecek2015} (see horizontal
lines in Fig.\,\ref{Figure4}). The total density of topological
defects for the nanocrystals, $\rho_{\rm def}=N_{\rm def}/N_{\rm
v}$, is always larger than $\rho_{\rm def}^{\rm macro}$. For a fixed
$B$, $\rho_{\rm def}$ enhances on increasing confinement. On the
other hand, for a fixed sample size $\rho_{\rm def}$ increases when
the structure softens on lowering $B$.

The abrupt increase of $\rho_{\rm def}(r)$ at the vicinity of the
sample edge occurs in a larger typical distance when increasing the
nanocrystal flexibility (decreasing $B$). We quantified this
tendency by
 fitting the data with a  $\rho_{\rm
def}(r)= A_{1}+ A_{2}\exp{[(r/a - d/2a)/\alpha]}$ dependence, see
full lines in Fig.\,\ref{Figure4}. The parameter $\alpha$ can be
interpreted as a number of lattice parameters in which the vortex
nanocrystal relaxes the shear stress induced by the bending of the
outer vortex-lines towards the center of the sample. This relaxation
is performed via the nucleation of plastic deformations. We will
thus regard $a\cdot\alpha$ as a characteristic ``healing-length''.
The evolution of $\alpha$ with $d$ for the three studied vortex
densities is shown in Fig.\,\ref{Figure4} (c). The length $\alpha$,
in lattice parameter units, increases with sample size. This
indicates that for larger nanocrystals the nucleation of topological
defects for mimicking the sample edge is performed more gradually
than for smaller nanocrystals. In addition, for a fixed sample size
$\alpha$ increases on softening the vortex nanocrystal.

The healing length $a \cdot \alpha$ described above may be an useful
quantity to model the $B$ and $d$-dependent entropy-jump depletion
in the vortex first-order transition. This depletion refers to
observation of a smaller $\Delta S / \Delta S_{0}$ at the
first-order melting transition of nanocrystalline vortex matter as
compared to the case of macroscopic samples, see insert to Fig.\,4
in Ref.\,\onlinecite{Dolz2015}.
 It is worth noting that the value $\alpha$
we detect is finite and no appreciable spatial gradient in vortex
density is observed. This is in sharp contrast to the effects
observed in other soft condensed matter systems such as
disk-confined one-component plasmas, where the density is
non-uniform and topological defects appear in the whole
system~\cite{mughal2007}. Nevertheless, this finding qualitatively
agrees with simulations for the case of systems with short-range
interactions \cite{lai1999} which are more adequate to model
vortex-vortex interactions in thick superconducting samples.
However, unlike these theoretical case-studies, vortices are line
objects that interact with bulk point disorder and in the case of
vortex nanocrystals they also interact with the sample surface that
align the outer-most vortices along its border and bend vortex rows.
This makes our problem much more complex. In particular: (a) the
expected equilibrium bulk order is not the crystalline Abrikosov
order; (b) the equilibration dynamics is expected to be glassy (much
slower than for a clean system) such that the equilibrium may not be
reached completely in a decoration experiment. One may thus ask
whether at distances from the sample edge larger than $a \cdot
\alpha$ the spatial order becomes the equilibrium one expected for
bulk disordered samples. In what follows we show evidence that this
seems to be the case, and that in the interior of samples, vortex
nanocrystals nucleate and are able to locally equilibrate  a
structure consistent with the Bragg-glass
phase.~\cite{Giamarchi1994}

In order to quantify the impact of pinning we will consider the
distance-evolution of the average displacement correlator  $B(r) =
\langle [u(r) - u(0)]^{2} \rangle/2$ evaluating the average over
quenched disorder and thermal fluctuations of the displacement of
vortices with respect to the sites of a perfect triangular lattice,
$u(r)$. In the case of macroscopic vortex matter, the theoretical
prediction~\cite{Giamarchi1994} states that the displacement
correlator presents three different regimes as a function of $r/a$.
Within the Larkin regime expected at short distances $B(r) \propto
r$  and the pinning is not yet effective in generating lattice
distorsions. When the displacements reach the scale in which pinning
is effective, i.e. when $[u(r) - u(0)] \sim \xi$, the system is in
the random manifold regime. In this regime the  evolution of the
displacement correlator is algebraic with distance, $B(r) \propto
r^{\nu}$ with $\nu=0.44$ for three-dimensional structures. In
addition, in this regime a constant ratio of
transversal-to-longitudinal displacement correlators is expected,
$B^{\rm T}(r)/B^{\rm L}(r) \sim 1.44$. The longitudinal correlation
function is defined as $B_{L}(r) = \langle {[\mathbf{u}(r) -
\mathbf{u}(0)]\cdot \mathbf{r}/r}^{2}\rangle /2$. The displacements
perpendicular to the vortex structure main directions are quantified
by the transverse displacement correlator that can be obtained as
$B_{T} = 2B(r) - B_{L}(r)$. For sufficiently large distances such
that $[u(r) - u(0)]
> a$ the Bragg-glass structure enters in the
quasi-ordered regime in which the dependence of $B(r)$ is
logarithmic with distance.

\begin{figure*}[!ht]
\begin{center}
\includegraphics[width=0.7\textwidth]{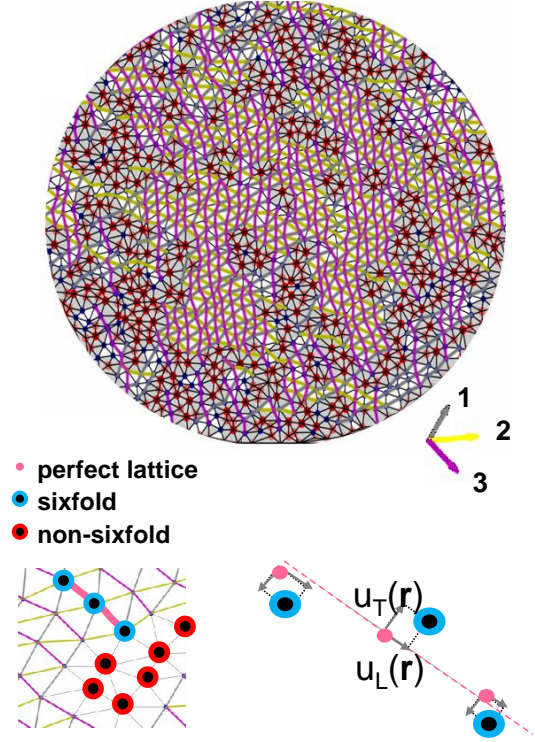}
\end{center}
\caption{\label{Figure6} Top: Lanes considered to calculate the
local vortex displacements $u(r)$ in the case of the 16\,G vortex
nanocrystal nucleated in the 50\,$\mu$m disk of Fig.\,\ref{Figure1}
(a). The lanes running parallel to one of the three principal
directions of the vortex structure are identified with the same
color.  Bottom: Schematics of the vortex displacements computed in
order to calculate the displacement correlator $B^{*}(r)$ along
($u_{\rm L}(r)$), and perpendicular to ($u_{\rm T}(r)$), a given
lane at a distance $r$ from its starting point. The real position of
vortices are indicated with large dots whereas the positions
corresponding to a perfect triangular lattice  are shown in small
gray dots.}
\end{figure*}

Direct imaging of vortex structures in dislocation-free regions of
Bi$_2$Sr$_2$CaCu$_2$O$_{8+\delta}$ vortex matter \cite{Kim1999}
showed the stabilization of the Larkin and random manifold regimes.
The Bragg glass regime was not directly observed since at distances
larger than 100\,$a$ the structure presents topological defects and
thus $u(r)$ is not well defined. In order to estimate $B(r)$ in our
vortex nanocrystals we implemented an algorithm for locally
calculating the displacement correlator in the presence of
topological defects. The algorithm calculates $B(r)$ not in the
whole vortex nanocrystal but in regions, considering lanes of
vortices in the three principal directions of the structure. The
regional lanes stop running two lattice parameters away of any
topological defect. The schematic representation of
Fig.\,\ref{Figure6} indicates how the lanes are defined in a given
vortex structure and how $B(r)$ is calculated for every lane. The
figure also illustrates the lanes (color lines) identified for
instance in a 16\,G vortex nanocrystal nucleated in a
50\,$\mu$m-diameter disk. The algorithm computes the
$B^{i}_{k}(r)$,with $k=1,2,3$ the three principal directions of the
structures, up to a distance equal to the length of every $i$-th
lane. Then we average the results obtained for all the $i-$th
parallel lanes within the $k$-direction in order to obtain $B_{k}$.
Finally we calculate $B^{*}$ by averaging the three $B_{k}$
magnitudes associated to the main directions.

For macroscopic Bi$_2$Sr$_2$CaCu$_2$O$_{8+\delta}$ vortex matter
with a density of 16\,G  we do not observe the Larkin regime
$B^{*}(r)\sim r$, indicating that the Larkin length is smaller than
our spatial resolution for $u(r)$. We do observe however that
$B(r)^{*} \sim (r/a)^{\nu}$ with $\nu = (0.4 \pm 0.2)$ and $B^{\rm
T}(r)/B^{\rm L}(r) \sim (1.4 \pm 0.5) $ up to the $r/a \sim
10$-range (see Fig.\ref{Figure6} (a)), as
 theoretically expected for an equilibrated random-manifold regime.
 In the case of the more
dense macroscopic vortex structures of 32 and 50\,G we find the same
exponent for the algebraic dependence of $B^{*}(r)$ within 5\,\%
dispersion for such a short distance-scale.
 Therefore, vortex structures in
macroscopic samples are in agreement with the nucleation of a
Bragg-glass phase. Since the decorated structure is expected to be a
snapshot of a configuration freezed at a characteristic temperature
$T_{\rm freez}\sim T_{\rm irr}$~\cite{Fasano2005,Bolecek2015}, our
results yield information on creep relaxation (random manifold)
dynamics at such a temperature. In particular, they show that at
$T_{\rm freez}$  length-scales as large as $\sim 10 a \sim
5-15$\,$\mu$m  can get equilibrated in the experimental time scale.

\begin{figure*}[!ht]
\begin{center}
\includegraphics[width=0.8\textwidth]{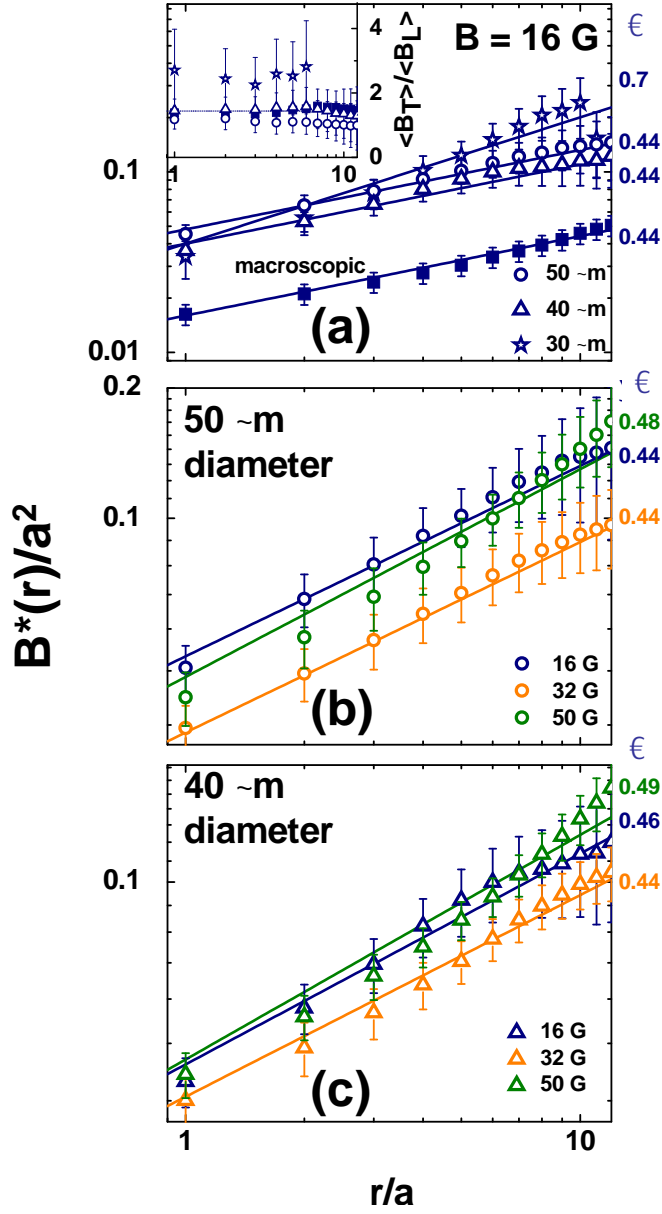}
\end{center}
\caption{\label{Figure7} Evolution of the normalized displacement
correlator  $B^{*}(r)/a^{2}$ with $r/a$  for vortex nanocrystals
nucleated in micron-sized Bi$_2$Sr$_2$CaCu$_2$O$_{8+\delta}$ disks.
(a) Data for 16\,G vortex structures nucleated in macroscopic
samples (full squares) and disks (open symbols). The insert shows
the transversal-to-longitudinal displacement correlator ratio and
dashed lines indicate the 1.44 value expected theoretically.
Magnetic field-evolution of the displacement correlator  for disks
of (b) 50 and (c) 40\,$\mu$m diameters. Full lines are  fits to the
data with an algebraic decay with an exponent $\nu$ indicated for
each case at the right.}
\end{figure*}

This scale of equilibration of the random-manifold regime is of the
order of the vortex nanocrystals we study. The displacement
correlator in nanocrystalline vortex matter, as a function of field
and sample physical size, is shown in Figs.\,\ref{Figure7} (b) and
(c). The absolute value of $B^{*}(r)/a^{2}$  for vortex nanocrystals
is more than 30\% larger  than for macroscopic samples. The $r/a$
evolution of the displacement correlator is algebraic even for the
smallest nanocrystal with roughly 400 vortices.

The exponent $\nu$ and the $B_{\rm T}/B_{L}$ ratio have values
expected for the random-manifold regime within the error, but in the
case of the smallest 16\,G nanocrystal both magnitudes increase
considerably. This can be due to the bending of vortex rows close to
the sample edge dominating the nucleation of the smallest studied
nanocrystal. Indeed, the Delaunay triangulation of
Fig.\,\ref{Figure1} (c) shows a structure with very small
crystallites. When increasing the vortex density for the same sample
size, the structure also presents an algebraic $B^{*}(r)$
 with an exponent of roughly $0.4$ within the error of $0.7$, whereas
 it slightly enhances on increasing vortex density, see
 Figs.\,\ref{Figure7} (b) and (c). In summary,
 except for those cases in which  more than 20\% of
 vortices are located at the nanocrystal surface and therefore
 confinement effects are dominant, the positional order of
 nanocrystalline vortex matter is also
 consistent with the nucleation of a
 Bragg-glass phase for distances from the edge towards the  center of the
 sample  larger than $\alpha$, equilibrated
 at lenght-scales belonging to its
 random-manifold regime of correlations.

The previous analysis shows that the relevance of bulk
 pinning for vortex nanocrystals with $400-1500$ vortices does
 not differ quantitatively from what is expected for macroscopic
 vortex matter. Physical properties that are affected by
 confinement in nanocrystalline vortex
 matter are thus controlled by the vortices located in
 a strip of width of order $a \cdot \alpha$ from the sample edges.
 The larger $a \cdot \alpha$ as compared to the sample diameter $d$,
 the more important are confinement effects.

\section{Conclusion}

We have shown that confinement affects the structure of vortex
nanocrystals nucleated in micron-sized
Bi$_2$Sr$_2$CaCu$_2$O$_{8+\delta}$ samples producing an excess of
topological defects within a characteristic healing length distance
from the edge. Towards the center of the nanocrystal, the vortex
lattice recovers the bulk-like structure observed in macroscopic
samples, consistent with the nucleation of a Bragg-glass phase. The
existence of a finite healing length may help to explain why
decreasing the system size to a few hundred vortices does not induce
any melting-point depression as observed for nanocrystals of hard
condensed matter,\cite{Goldstein1992} but entails a progressive
depletion of the entropy-jump at the first-order
transition.~\cite{Dolz2015} We recall that the entropy-jump
depletion can have two extra contributions rather than this
structural one. The first one can come from a change on the coupling
of pancakes making up the outer-most vortex lines when reducing the
nanocrystal size. The second one can come from the possibility of
the sample surface straightening the outer-most vortices as any
correlated disorder would produce. Further experimental and
theoretical investigations on the magnetic field, temperature and
sample size-dependence of the healing length would be thus promising
in order to build a simple phenomenological model for the anomalous
properties of small confined systems of vortices and alike.

\section{Acknowledgments}

We thank Ming Li for sample preparation.

\end{document}